\documentclass[conference]{IEEEtran}
\usepackage{cite}
\ifCLASSINFOpdf
   \usepackage[pdftex]{graphicx}
   \graphicspath{{./pdf/}{./jpeg/}{./png/}}
   \DeclareGraphicsExtensions{.pdf,.jpeg,.png,.jpg}
\else
   \usepackage[dvips]{graphicx}
   \graphicspath{{../eps/}}
   \DeclareGraphicsExtensions{.eps}
\fi
\usepackage[acronym, nomain,nogroupskip,nonumberlist,nopostdot, toc]{glossaries}
\usepackage{pbox}

\newacronym{4G}{4G}{fourth generation}
\newacronym{ADC}{ADC}{Analogue-to-Digital Converter}
\newacronym{ADS}{ADS}{Advanced Design System}
\newacronym{AoA}{AoA}{Angle of Arrival}
\newacronym{AoD}{AoD}{Angle of Departure}
\newacronym{AP}{AP}{Access Point}
\newacronym{AS}{AS}{angular spread}
\newacronym{AWGN}{AWGN}{Additive White Gaussian Noise}
\newacronym{BBU}{BBU}{Base Band Unit}
\newacronym{BCC}{BCC}{Bristol City Council}
\newacronym{BER}{BER}{Bit Error Rate}
\newacronym{BF}{BF}{Beamforming}
\newacronym{BIO}{BIO}{Bristol Is Open}
\newacronym{BS}{BS}{Base Station}
\newacronym{BW}{BW}{Bandwidth}
\newacronym{CCDF}{CCDF}{complementary cumulative distribution function}
\newacronym{CDF}{CDF}{Cumulative Distribution Function}
\newacronym{CDMA}{CDMA}{Code Division Multiple Access}
\newacronym{CDT}{CDT}{Centre for Doctoral Training}
\newacronym{COTS}{COTS}{Commercial off-the-shelf}
\newacronym{CP}{CP}{Cyclic Prefix}
\newacronym{CPA}{CPA}{coordinated pilot assignment}
\newacronym{CRC}{CRC}{Cyclic Redundancy Check}
\newacronym{CSI}{CSI}{Channel State Information}
\newacronym{CSIRO}{CSIRO}{Commonwealth Scientific and Industrial Research Organisation}
\newacronym{CSN}{CSN}{Communication Systems \& Networks}
\newacronym{D2D}{D2D}{device-to-device}
\newacronym{DAC}{DAC}{Digital-to-Analogue Converter}
\newacronym{dB}{dB}{decibels}
\newacronym{DL}{DL}{downlink}
\newacronym{DMA}{DMA}{Direct Memory Access}
\newacronym{DSP}{DSP}{Digital Signal Processing}
\newacronym{EM}{EM}{Electro-Magnetic}
\newacronym{EPSRC}{EPSRC}{Engineering and Physical Sciences Research Council}
\newacronym{ESD}{ESD}{Electro-Static Discharge}
\newacronym{FFT}{FFT}{Fast Fourier Transform}
\newacronym{FDD}{FDD}{Frequency Division Duplex}
\newacronym{FIFO}{FIFO}{First-in First-out}
\newacronym{FITRA}{FITRA}{Fast Iterative Truncation Algorithm}
\newacronym{FOC}{FOC}{Frequency Offset Correction}
\newacronym{FPGA}{FPGA}{Field-Programmable Gate Array}
\newacronym{FXP}{FXP}{fixed-point}
\newacronym{GPS}{GPS}{Global Positioning System}
\newacronym{GPSDO}{GPSDO}{\gls{GPS} Disciplined Oscillator}
\newacronym{GNSS}{GNSS}{Global Navigation Satellite Systems}
\newacronym{H2T}{H2T}{Host-to-target}
\newacronym{HD}{HD}{High-Definition}
\newacronym{HDL}{HDL}{Hardware Description Language}
\newacronym{ICT}{ICT}{Information and Communications Technology}
\newacronym{ID}{ID}{Identity}
\newacronym{IFFT}{IFFT}{Inverse Fast Fourier Transform}
\newacronym{iid}{iid}{independent and indentically distributed}
\newacronym{IO}{IO}{Input/Output}
\newacronym{IP}{IP}{Intellectual Property}
\newacronym{ISI}{ISI}{Inter-Symbol Interference}
\newacronym{JSDM}{JSDM}{Joint Spatial Division Multiplexing}
\newacronym{LDPC}{LDPC}{Low-Density Parity-Check}
\newacronym{LED}{LED}{Light Emitting Diode}
\newacronym{LF}{LF}{Loading Factor}
\newacronym{LIDAR}{LIDAR}{Laser Illuminated Detection \& Ranging}
\newacronym{LO}{LO}{Local Oscillator}
\newacronym{LOS}{LOS}{line-of-sight}
\newacronym{LS}{LS}{least squares}
\newacronym{LSD}{LSD}{Local Search Detection}
\newacronym{LTE}{LTE}{Long-Term Evolution}
\newacronym{LUT}{LUT}{Look-up Table}
\newacronym{MAC}{MAC}{Medium Access Control}
\newacronym{MF}{MF}{Matched Filtering}
\newacronym{MRT}{MRT}{Maximal Ratio Transmission}
\newacronym{MCS}{MCS}{Modulation and Coding Scheme}
\newacronym{MGS}{MGS}{Modified Gram Schmidt}
\newacronym{MIMO}{MIMO}{Multiple-Input, Multiple-Output}
\newacronym{ML}{ML}{Maximum likelihood}
\newacronym{MSE}{MSE}{Mean Square Error}
\newacronym{MMSE}{MMSE}{Minimum Mean Square Error}
\newacronym{MPD}{MPD}{Message Passing Detection}
\newacronym{MRC}{MRC}{Maximal Ratio Combining}
\newacronym{MU}{MU}{Multi-user}
\newacronym{MUI}{MUI}{Multi-User Interference}
\newacronym{MUSIC}{MUSIC}{Multiple Signal Classification}
\newacronym{MXI}{MXI}{Multisystem Extension Interface}
\newacronym{NI}{NI}{National Instruments}
\newacronym{NLOS}{NLOS}{Non-line-of-sight}
\newacronym{OBR}{OBR}{Out-of-band power ratio}
\newacronym{OCXO}{OCXO}{Oven-controlled Crystal Oscillator}
\newacronym{OFDM}{OFDM}{Orthogonal Frequency Division Multiplexing}
\newacronym{OTA}{OTA}{over-the-air}
\newacronym{PA}{PA}{Power Amplifier}
\newacronym{PAPR}{PAPR}{peak-to-average power ratio}
\newacronym{PAS}{PAS}{Power-Azimuth Spectrum}
\newacronym{PC}{PC}{Pilot Contamination}
\newacronym{PCIe}{PCIe}{Peripheral Component Interconnect Express}
\newacronym{PDP}{PDP}{Power Delay Profile}
\newacronym{PDSCH}{PDSCH}{Physical Downlink Shared Channel}
\newacronym{PHY}{PHY}{Physical Layer}
\newacronym{PL}{PL}{Path-Loss}
\newacronym{PLL}{PLL}{Phase-locked Loop}
\newacronym{PMP}{PMP}{precoding, modulation and \gls{PAPR} reduction}
\newacronym{P2P}{P2P}{Peer-to-peer}
\newacronym{ppb}{ppb}{parts-per-billion}
\newacronym{PPC}{PPC}{Pilot Power Control}
\newacronym{PSS}{PSS}{Primary Synchronisation Signal}
\newacronym{PTS}{PTS}{Pilot Time Slot}
\newacronym{PUSCH}{PUSCH}{Physical Uplink Shared Channel}
\newacronym{PXIe}{PXIe}{\gls{PCIe} eXtensions for Instrumentation}
\newacronym{QAM}{QAM}{Quadrature Amplitude Modulation}
\newacronym{QPSK}{QPSK}{Quadrature Phase-Shift Keying}
\newacronym{QRD}{QRD}{QR Decomposition}
\newacronym{RAN}{RAN}{Radio Access Network}
\newacronym{RB}{RB}{Resource Block}
\newacronym{R&D}{R\&D}{Research and development}
\newacronym{RE}{RE}{Resource Element}
\newacronym{RF}{RF}{Radio Frequency}
\newacronym{RIO}{RIO}{Reconfigurable Input/Output}
\newacronym{RRH}{RRH}{Remote Radio Head}
\newacronym{RSS}{RSS}{Received Signal Strength}
\newacronym{RT}{RT}{Ray-tracing}
\newacronym{RTM}{RTM}{Rear Transition Module}
\newacronym{RX}{RX}{Receive}
\newacronym{SC}{SC}{Single-Carrier}
\newacronym{SDMA}{SDMA}{Space-Divison Multiple Access}
\newacronym{SDR}{SDR}{Software-Defined Radio}
\newacronym{SER}{SER}{Symbol Error Rate}
\newacronym{SINR}{SINR}{Signal to Interference plus Noise Ratio}
\newacronym{SIR}{SIR}{Signal to Interference Ratio}
\newacronym{SM}{SM}{Spatial Modulation}
\newacronym{SMA}{SMA}{SubMiniature version A}
\newacronym{SNR}{SNR}{Signal to Noise Ratio}
\newacronym{SPDT}{SPDT}{Single Pole Double Throw}
\newacronym{SISO}{SISO}{Single-Input Single-Output}
\newacronym{SVD}{SVD}{Singular Value Decomposition}
\newacronym{SVS}{SVS}{Singular Value Spread}
\newacronym{T2H}{T2H}{Target-to-host}
\newacronym{TDD}{TDD}{Time Division Duplex}
\newacronym{TDOA}{TDOA}{Time-Difference-of-Arrival}
\newacronym{TO}{TO}{Timing Offset}
\newacronym{TR}{TR}{tone-reservation}
\newacronym{TX}{TX}{Transmit}
\newacronym{TTI}{TTI}{Transmission Time Interval}
\newacronym{U8}{U8}{Unsigned 8-bit}
\newacronym{UDP}{UDP}{User Datagram Protocol}
\newacronym{UE}{UE}{User Equipment}
\newacronym{UI}{UI}{User Interface}
\newacronym{UL}{UL}{Uplink}
\newacronym{USRP}{USRP}{Universal Software Radio Peripheral}
\newacronym{UoB}{UoB}{University of Bristol}
\newacronym{VLC}{VLC}{VideoLAN Client}
\newacronym{VTC}{VTC}{Vehicular Technology Conference}
\newacronym{WARP}{WARP}{Wireless Open-Access Research Platform}
\newacronym{ZF}{ZF}{Zero-Forcing}

\begin{document}
\title{An Overview of Massive MIMO Research at the University of Bristol}

%\author{\IEEEauthorblockN{Paul Harris}
%\IEEEauthorblockA{Communication Systems\\ \& Networks Group\\
%University of Bristol\\
%Bristol, UK\\
%Email: paul.harris@bristol.ac.uk}
%\and
%\IEEEauthorblockN{Siming Zhang}
%\IEEEauthorblockA{Communication Systems\\ \& Networks Group\\
%University of Bristol\\
%Bristol, UK\\
%Email: paul.harris@bristol.ac.uk}
%\and
%\IEEEauthorblockN{James Kirk\\ and Montgomery Scott}
%\IEEEauthorblockA{Starfleet Academy\\
%San Francisco, California 96678--2391\\


\author{\IEEEauthorblockN{Paul Harris,
Wael Boukley Hasan,
Henry Brice,
Benny Chitambira,
Mark Beach,
Evangelos Mellios\\
Andrew Nix,
Simon Armour,
Angela Doufexi\\
\IEEEauthorblockA{Communication Systems \& Networks Group,
University of Bristol,
Bristol, UK \\ Email: \{paul.harris, wb14488, henry.brice, b.chitambira, m.a.beach, evangelos.mellios\\ andy.nix, simon.armour, angela.doufexi\}@bristol.ac.uk}}}


\maketitle

\begin{abstract}
Massive \gls{MIMO} has rapidly gained popularity as a technology crucial to the capacity advances required for 5G wireless systems.	Since its theoretical conception six years ago, research activity has grown exponentially, and there is now a developing industrial interest to commercialise the technology. For this to happen effectively, we believe it is crucial that further pragmatic research is conducted with a view to establish how reality differs from theoretical ideals. This paper presents an overview of the massive \gls{MIMO} research activities occurring within the Communication Systems \& Networks Group at the University of Bristol centred around our 128-antenna real-time testbed, which has been developed through the \gls{BIO} programmable city initiative in collaboration with \gls{NI} and Lund University. Through recent preliminary trials, we achieved a world first spectral efficiency of 79.4 bits/s/Hz, and subsequently demonstrated that this could be increased to 145.6 bits/s/Hz. We provide a summary of this work here along with some of our ongoing research directions such as large-scale array wave-front analysis, optimised power control and localisation techniques.\\
\end{abstract}

% \fi
\IEEEpeerreviewmaketitle

\section{Introduction}
\gls{MIMO} has become a mature communications technology in recent years, finding itself incorporated today within both Wi-Fi and \gls{4G} cellular standards. Current systems typically deploy between 2 to 4 antennas at the \gls{AP} or \gls{BS}, and they can be used to either enhance the achievable throughput for a single device or allow 2 to 4 devices to be served simultaneously in the same frequency resource. Massive \gls{MIMO} takes the latter \gls{MU} \gls{MIMO} concept one step further by deploying hundreds of antennas at the \gls{BS}, each with their own individual \gls{RF} chain. The result is greatly enhanced spatial multiplexing performance allowing many tens of \glspl{UE} to be served with greater reliability than in standard \gls{MU} \gls{MIMO}. It is well-recognised as one of the key enabling technologies for 5G that could provide superior spectral and energy efficiencies. The theoretical benefits of massive \gls{MIMO} can be found discussed and well documented in \cite{Marzetta2010},\cite{Hoydis2013} and \cite{6736761}.\\

In this paper, we provide an overview of the massive MIMO research activities occurring with the Communication Systems \& Networks Group at the University of Bristol, centred round our 128-antenna real-time testbed. In addition to preliminary results from these first measurement trials, we highlight some of our key areas of interest, including optimised power control algorithms, client localisation and wave front analysis through ray tracing models.



\section{A Pragmatic Focus}

The theoretical advantages for moving towards such an extreme case of \gls{MU} \gls{MIMO} have been widely published over recent years and both academia and industry are now rapidly shifting their focus towards real-world tests. The \gls{BIO} massive \gls{MIMO} research platform, previously introduced in \cite{Harris2015}, has been developed within the \gls{CSN} Research Group at the University of Bristol in close collaboration with both National Instruments and Lund University, and it has begun to enable a range of pragmatic massive \gls{MIMO} research and world first results\cite{WorldRecord}\cite{WorldRecord2}.
This section will provide a light overview of our 128-antenna testbed and the two first indoor measurement trials conducted at the University of Bristol.

\begin{figure}[!t]
	\centering
	\includegraphics[width=\columnwidth]{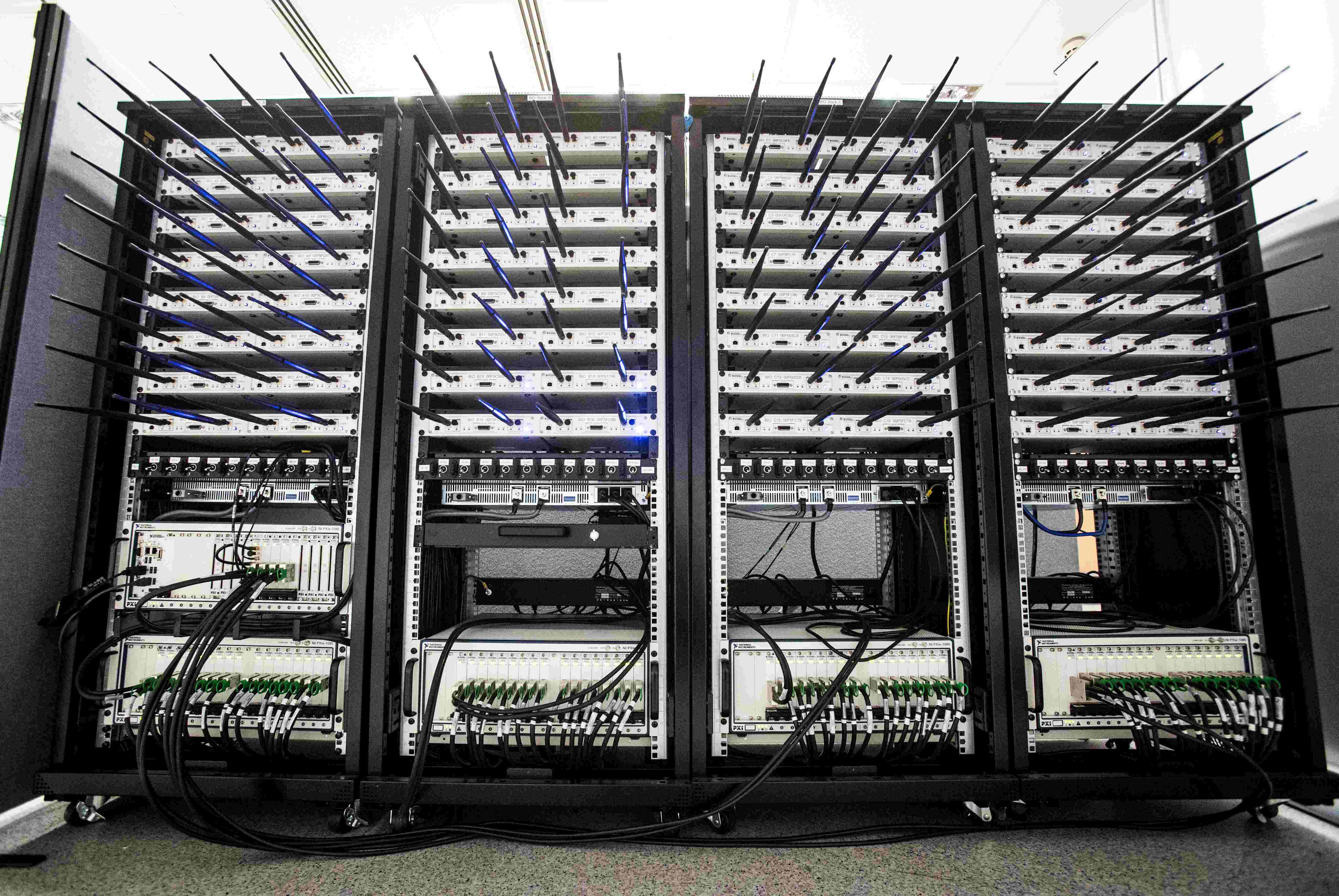}
	\caption{The \gls{BIO} Testbed}
	\label{fig_BS}
\end{figure}

\subsection{System Overview}
The \gls{BIO} massive \gls{MIMO} \gls{BS} shown in ~\figurename~\ref{fig_BS} consists of 64 \gls{NI} \gls{USRP} \gls{RIO} \cite{USRP} \glspl{SDR} providing 128 \gls{RF} chains, with a further 6 \gls{USRP} \glspl{RIO} acting as 12 single-antenna \glspl{UE}. It runs with an LTE-like \gls{PHY} and the key system parameters can be seen in Table \ref{tab:system_param}.

\begin{table}
	\renewcommand{\arraystretch}{1.3}
	\caption{System Parameters}
	\centering
	\noindent\begin{tabular}{ll}
		% \toprule
		%\begin{tabular}{|l|l|l|}
		\textbf{Parameter} & \textbf{Value}  \\
		%\midrule
		\# of BS Antennas & 128  \\
		\# of UEs & 12 \\
		Carrier Frequency & 1.2-6 GHz (3.51 GHz licensed) \\
		Bandwidth		  & 20 MHz  \\
		Sampling Frequency  & 30.72 MS/s \\
		Subcarrier Spacing  & 15 kHz \\
		\# of Subcarriers  &  2048 \\
		\# of Occupied Subcarriers  & 1200 \\
		Frame duration  & 10 ms \\
		Subframe duration & 1 ms \\
		Slot duration & 0.5 ms \\
		TDD periodicity &   1 slot  \\	
	\end{tabular}
	\label{tab:system_param}
\end{table}

Using the \gls{NI} \gls{PXIe} platform, all the \glspl{RRH} and \gls{MIMO} \gls{FPGA} processors in the system are linked together by a dense network of gen 3 \gls{PCIe} fabric, and all software and \gls{FPGA} behaviour is programmed via LabVIEW. Further detail about the system architecture and the implementation of our wide data-path \gls{MMSE} encoder/decoder can be found in  \cite{PaulGlobecom} and \cite{PaulSIPS}.

\subsection{Initial Trials}

\subsubsection{Trial one}

The length of the lower atrium in the University of Bristol’s Merchant Venturers Building was used for three different \gls{LOS} measurements between the \gls{BS} and 12 \glspl{UE}. \glspl{UE} were grouped both in a straight line parallel to the \gls{BS} and at a slant, with a distance of 3.3m, 12.5m or 18.1m to the nearest client in each scenario. At the \gls{BS} side, a 5.44m 128-element linear array of dipoles was used, providing half-wavelength spacing at 3.5 GHz. In addition to capturing channel data for offline analysis, we managed to achieve a real-time uncoded sum-rate of 1.59 Gbps in only 20 MHz of \gls{BW}, equating to a record spectral efficiency of 79.4 bits/s/Hz \cite{WorldRecord}.

\subsubsection{Trial two}

\begin{figure}[!t]
	\centering
	\includegraphics[width=\columnwidth]{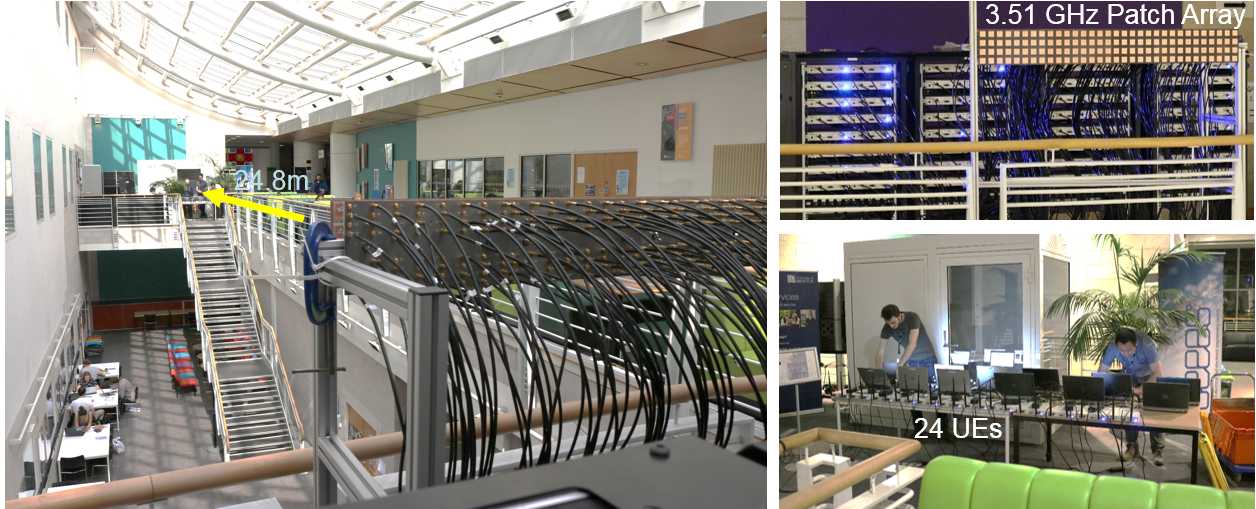}
	\caption{Second measurement trial with the \glspl{UE} 24.8m away}
	\label{fig_trial2}
\end{figure}

For the second trial, the upper level of the Merchant Venturers Building atrium was used with a patch panel antenna array to serve user clients placed 24.8m away on the opposite balcony. The array was setup in a 4x32 configuration with alternate H \& V polarisations for all 128 antennas. Following a code modification and the provisioning of additional client radios, we were able to perform decimated channel captures and host-based massive \gls{MIMO} detection for up to 24 users, allowing us to observe recovered constellations and channel statistics in real-time. As with the first trial, the \glspl{UE} were in \gls{LOS} and placed in a straight line with 2.5 $\lambda$ spacing. However, this environment was not so static, as it was a normal working day and students were present. An overview of the setup can be seen in ~\figurename~\ref{fig_trial2}.

In the scenario described, we were able to recover clear 256-QAM \gls{UL} constellations for 22 users. Using the same frame schedule as in trial one, this would scale the achievable real-time throughput and spectral efficiency to nearly 3 Gbps and 145.6 bits/s/Hz respectively \cite{WorldRecord2}. More can be found out about both trials in \cite{PaulGlobecom} and \cite{PaulSIPS}.

\section{Channel \& Wavefront Analysis}
This section presents an outline of methods that can be used to both analyse and model the propagation characteristics of the massive \gls{MIMO} channel. Although many of the characteristics inherent in standard \gls{MIMO} channels are also present in massive \gls{MIMO} channels, there are some significant differences, such as the need to consider spherical wavefronts rather than plane wave ones \cite{zhou2015spherical} and the presence of slow fading across large arrays \cite{aulin2015benefits}. The measurement campaigns that have been conducted at the University of Bristol may also be able to reveal other important characteristics of the channel that have not been widely documented as well as providing clarity with regard to known phenomena. Some of the methods that have been used to analyse the characteristics are discussed here. This is followed by an overview of propagation modelling techniques, in particular the ray-tracing system that can be used to enable a more detailed analysis in tandem with the outdoor measurements, followed by a description of how different types of wave front models can be used to approximate the channel.

\subsection{Fading Across the Array}
\begin{figure}[!t]
	\centering
	\includegraphics[width=\columnwidth]{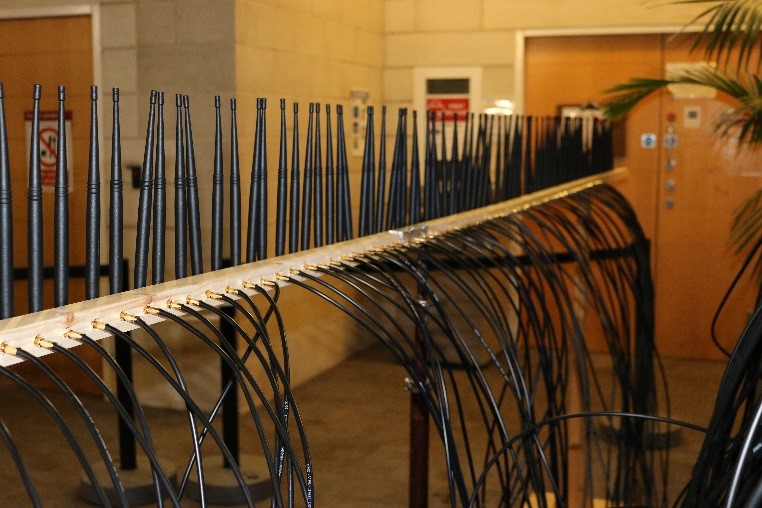}
	\caption{5.4m Linear Array of Dipoles used in initial indoor experiment}
	\label{fig_linear}
\end{figure}

\begin{figure}[!t]
	\centering
	\includegraphics[width=\columnwidth]{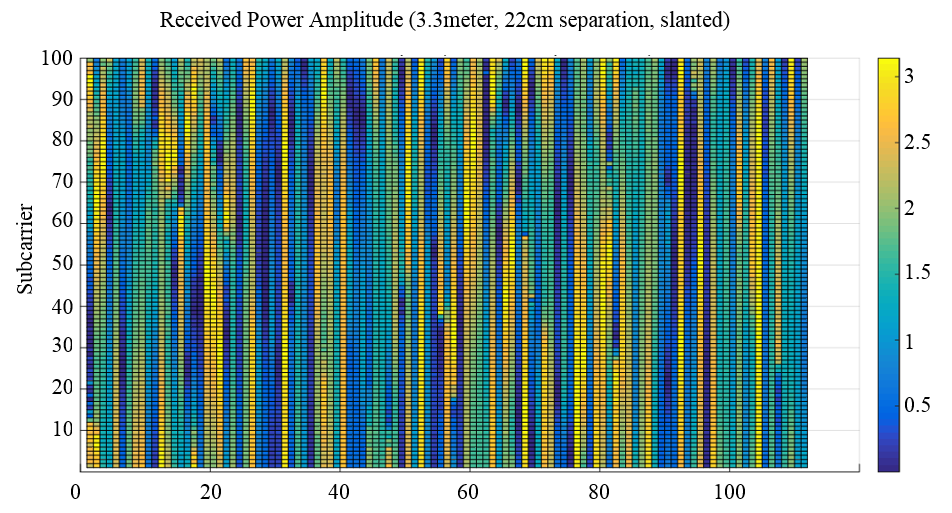}
	\caption{Normalised power received at each \gls{BS} antenna for one user}
	\label{fig_power}
\end{figure}

The use of the linear array shown in~\figurename~\ref{fig_linear} allows for the observation of large-scale changes. The testbed periodically captures a snapshot of the full channel frequency response between all 128 \gls{BS} antennas and the 12 single-antenna users for all 1200 \gls{OFDM} subcarriers, with a resolution of one resource block (12 subcarriers). This resolution results from the use of frequency-orthogonal pilots with each user transmitting on every twelfth 15 kHz subcarrier originating at its user ID (1-12). The normalised power received at each base station antenna can then be obtained by considering a signal of unity power transmitted from each of the mobile stations, as shown in ~\figurename~\ref{fig_power}. Statistical techniques can then be used to extract relevant information that can be compared with the propagation models.

\subsection{Coherence Bandwidth / Delay Spread}
With the measured data, it is possible to obtain not only the coherence \gls{BW} between two antennas, which is a key parameter for any channel when designing a network, but also to be able to determine how this parameter changes from the perspective of a mobile antenna as it ‘looks’ across the array from one side to the other. This is possible because, even though the raw data does not include the sampled impulses for each of the \gls{OFDM} subcarriers, it is possible to recover all the relevant sample points by using standard techniques for digital-to-analogue conversion, but in the frequency domain. The time-domain impulse response (which is the power-delay profile) can then be recovered using the \gls{IFFT}, since it is related directly to the frequency domain impulse response.

\subsection{Ray Tracing}
The ray-tracing system developed by the University of Bristol allows for the obtaining of a channel impulse response for any transmitter and receiver location by making use of reflection, transmission and refraction \cite{athanasiadou1995ray}. The rays are also calculated in three-dimensions, allowing for the incorporation of antenna patterns. A database of the city of Bristol is available that makes it possible to develop a deterministic model of outdoor measurement campaigns such as the one that was conducted recently near to the Merchant Ventures Building. The ray-tracing system is especially useful for massive \gls{MIMO} campaigns using a large linear array because it enables experimenters to investigate in detail the changes across the array, and in particular how effects such as slow fading (resulting, for example, from part of the array being shadowed by a building whilst the other part has a \gls{LOS} link with the mobile user) affect the channel as a whole.

\subsection{Spherical \& Planar Models}
By extending \cite{bohagen2006modeling}, it is possible to model the channel matrix as
\begin{equation}
\mathbf{H} = \mathbf{a}_{r}\mathbf{a}_{t}^{T}
\end{equation}
where $\mathbf{a}_{r}$ and $\mathbf{a}_{t}$ are the spatial signatures for the receiver and transmitter respectively, obtained by making a plane wave assumption across the entire transmit and receive arrays while conveniently not considering each individual antenna element.

Experimental research has shown that a standard planar wave model like this is often inadequate for large linear arrays because of its inability to correctly model the line-of-sight component \cite{jiang2005spherical}. This necessitates the use of a spherical wave model that requires a computation of the wave front between each of the antennas such as
\begin{equation}
\mathbf{H}_{m,n} = e^{j\frac{2\pi}{\lambda}\mathbf{r}_{m,n}}
\end{equation}
where $\mathbf{r}_{m,n}$ is the distance between each transmit and receive antenna, denoted by the subscripts $m$ and $n$. This inevitably leads to an increase in computational complexity. It is possible to use the measured data to not only verify the validity of spherical models, but also to identify scenarios where less computational costly approaches could be used instead.

\section{Power Control}

The right power control in a massive \gls{MIMO} system will help improve terminal \glspl{SINR} and increase the performance for users at the cell edge. Like \gls{CDMA} systems, it is crucial for mitigating the near-far problem and ensuring balanced performance. In massive \gls{MIMO}, the ‘channel hardening’ phenomenon that results from using a large number of antennas at the BS opens up new possibilities for the efficient implementation of such algorithms \cite{Marzetta2010}. The channel hardening in massive \gls{MIMO} was discussed and pictorially illustrated in \cite{narasimhan2014channel}.

\begin{figure}[!t]
	\centering
	\begin{minipage}{0.49\columnwidth}
		\centering
		\includegraphics[width=\textwidth]{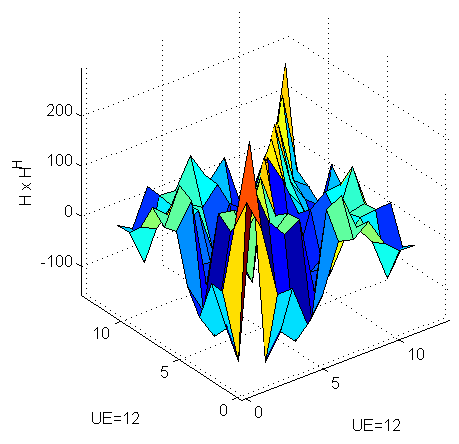}
	\end{minipage}\hfill
	\begin{minipage}{0.49\columnwidth}
		\centering
		\includegraphics[width=\textwidth]{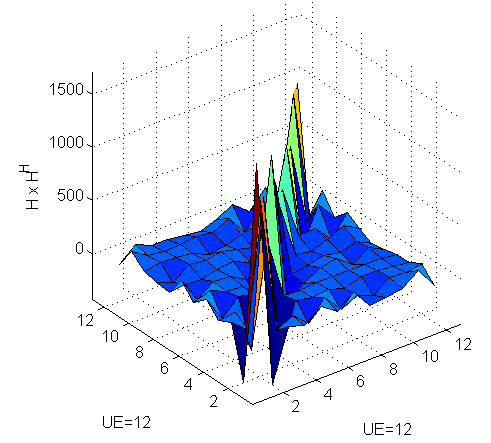}
	\end{minipage}
	%\centering
	
	\caption{$\mathbf{HH}^{H}$ Channel Gram Matrix (not normalised). Left: 32 antennas at the BS. Right: 112 antennas at the BS. 12 single-antenna \glspl{UE} in each case.}
	\label{Gram}
\end{figure}

In the aforementioned measurement trials, real time channel measurements were captured by the massive \gls{MIMO} testbed and twelve single antenna clients were supported. Before we designed a power control algorithm for massive \gls{MIMO}, two experiments took place within these trials to investigate the practicality of relying on the channel hardening in our design. In the first experiment, the channel was measured by the massive \gls{MIMO} testbed with only 32 active \gls{BS} antennas and the distance between the \gls{BS} and the clients was 20.81 m. ~\figurename~\ref{Gram} left shows the channel gram matrix for a static indoor environment averaged over 100 captures using 32 \gls{BS} antennas. On the second experiment, the number of active antennas at the \gls{BS} was increased to 112, and the distance between the \gls{BS} and the clients was reduced to 11.6 m. This time, the measurement environment was not static since three people walked randomly between the \gls{BS} and the clients during the channel measurements, and data was averaged over 400 captures. ~\figurename~\ref{Gram} right illustrates experiment two when 112 active \gls{BS} antennas were used and the composite channel has clearly become more deterministic. By comparing the results from both experiments, the ratio between the eigenvalues and the maximum off-diagonal elements of the gram matrix was decreased from 44\% into 16\%.

Based upon the results obtained from these initial experiments, we designed an uplink power control algorithm for massive MIMO which exploits the channel hardening properties to perform closed loop control at the \gls{PHY}. The aim of our design is to increase the average \gls{SINR} and power efficiency whilst simultaneously decreasing the transmission overheads, latency and complexity of the receiver. This design was subsequently implemented and tested on the \gls{BIO} massive \gls{MIMO} testbed along with two additional uplink power control algorithms for comparison purposes. The first one is based on a constant \gls{SNR} value, whilst the second one is based on a constant \gls{SINR} value. Following the previous two trials, another indoor experiment took place and the power control was tested in real-time. The environment was changing during channel measurements, although the client devices remained static. The aggregate average \gls{SINR} was 5.4 \gls{dB} when the power control level adjustment was based on a fixed \gls{SNR} value. This value was increased by 0.4 when the power level adjustment was based on the \gls{SINR}. With the power control algorithm we designed, the aggregate average \gls{SINR} was enhanced by 0.5 dB compared to when a fixed \gls{SINR} was used.      

%\begin{figure}[!t]
%	\centering
%	\includegraphics[width=2in]{SumrateDisplay}
%	\caption{Real-time Display of system Sum-rate with 3.3m UE group distance and 22cm client separation}
%	\label{fig_sumrate}
%\end{figure}

\section{Localisation}
Accurate geolocation in urban environments is a challenge. \gls{GNSS} require \gls{LOS} communication with at least 3 satellites, which can be difficult in cities due to the urban canyon \cite{GSA}. Massive \gls{MIMO} represents an opportunity for mobile radio network based localisation because inexpensive, low-power and low-precision components can be used, with greatly reduced complexity and cost in terms of antenna requirements and equipment calibration.

Other strong motivations for considering mobile network localisation with massive \gls{MIMO} are the significant potential benefits this brings to a massive \gls{MIMO} system itself, and possibly other next-generation wireless systems like mmWave. If a \gls{BS} could build a picture of the way mobile devices are moving through the environment, it would potentially give rise to the following:
\begin{itemize}
	\item New handover strategies - By using geolocation information together with inertial measurements, a desirable handover point could be predicted.
	\item New resource management strategies - A mobile position can be compared to a posteriori information from heat-map style based tools to implement appropriate adaptive modulation and coding schemes.
	\item Power Control and reduction in device transmit power - When a mobile device moves from a \gls{LOS} position to a highly shadowed, \gls{NLOS} position, the change is likely to be abrupt, and closed-loop power control algorithms may struggle. This would lead to power control errors in such a scenario. If power control algorithms could use the location information, together with knowledge of the environment, power control errors could be reduced. Massive \gls{MIMO} detection is also very robust compared to \gls{SISO} systems and devices may be allowed to transmit at the minimum levels.
	\item Reduce pilot contamination in dense deployments - Location-based channel estimation improves the overall system performance. Pilot allocation can be made such that all mobiles with similar \glspl{AoA} are prevented from sharing the same pilot \cite{WangLocation}.
	\item Geolocation information can also be used for mmWave beamforming - Localisation can be performed using massive \gls{MIMO} at sub 6 GHz in the mobile network, but the geolocation information can then be used to adjust the downlink beamforming for mmWave.
\end{itemize}

Theory and simulations demonstrate that super resolution schemes like the \gls{MUSIC} perform better as the number of antenna elements is increased. An array of 100 elements should produce very sharp peaks in the \gls{PAS}, which makes \gls{AoA} or \gls{AoD} estimation in massive \gls{MIMO} very reliable. Furthermore, rectangular arrays would also make elevation \gls{AoA}/\gls{AoD} estimation possible.
Using the \gls{BIO} testbed, the performance and limitations of massive \gls{MIMO} for localisation can be explored. Single \gls{BS} localisation is possible in scenarios where the mobile device is known to be in \gls{LOS}, but distribution opens up more opportunities. Due to the modular nature of the testbed, it is distributable into subarrays, and each subsystem can still be synchronised to a common clock. Techniques such as \gls{TDOA} can then be utilised for mobile clients seeing at least 3 of these distributed arrays.

Most of the potential system benefits identified herein depend on the \glspl{BS} building a picture of the environment around them. \gls{LOS} identification for signals between the \gls{BS} and a mobile client is therefore very critical. \gls{LOS} and \gls{NLOS} identification is also a key feature of most localisation algorithms, and for these reasons, identification techniques have been developed and tested using ray tracing simulations based on a real world \gls{LIDAR} database of the city of Bristol.

\section{Ongoing and Future Work}

Future work will include real-time \gls{DL} performance evaluation, implementation of massive \gls{MIMO} optimised power control, \gls{OTA} synchronisation optimisation, rooftop deployments and node distribution on the \gls{BIO} citywide fibre network. The massive \gls{MIMO} performance will be investigated with different kinds of mobility and suitable power control update rates will be determined for different operational scenarios.​

\section*{Acknowledgement}
The authors wish to thank Karl Nieman and Nikhil Kundargi from National Instruments for their ongoing support with the software development; professors Edfors, Tufvesson, PhD student Steffen Malkowsky and the rest of the Lund University research team for their contributions; and Bristol Is Open for access to the hardware facility. They also acknowledge the financial support of the \gls{EPSRC} \gls{CDT} in Communications (EP/I028153/1), NEC and \gls{NI}.



\bibliographystyle{IEEEtran}
\bibliography{library}

\begin{thebibliography}{10}
\providecommand{\url}[1]{#1}
\csname url@samestyle\endcsname
\providecommand{\newblock}{\relax}
\providecommand{\bibinfo}[2]{#2}
\providecommand{\BIBentrySTDinterwordspacing}{\spaceskip=0pt\relax}
\providecommand{\BIBentryALTinterwordstretchfactor}{4}
\providecommand{\BIBentryALTinterwordspacing}{\spaceskip=\fontdimen2\font plus
\BIBentryALTinterwordstretchfactor\fontdimen3\font minus
  \fontdimen4\font\relax}
\providecommand{\BIBforeignlanguage}[2]{{%
\expandafter\ifx\csname l@#1\endcsname\relax
\typeout{** WARNING: IEEEtran.bst: No hyphenation pattern has been}%
\typeout{** loaded for the language `#1'. Using the pattern for}%
\typeout{** the default language instead.}%
\else
\language=\csname l@#1\endcsname
\fi
#2}}
\providecommand{\BIBdecl}{\relax}
\BIBdecl

\bibitem{Marzetta2010}
T.~L. Marzetta, ``{Noncooperative Cellular Wireless with Unlimited Numbers of
  Base Station Antennas},'' \emph{IEEE Transactions on Wireless
  Communications}, vol.~9, no.~11, pp. 3590--3600, nov 2010.

\bibitem{Hoydis2013}
J.~Hoydis \emph{et~al.}, ``{Massive MIMO in the UL/DL of Cellular Networks: How
  Many Antennas Do We Need?}'' \emph{IEEE Journal on Selected Areas in
  Communications}, vol.~31, no.~2, pp. 160--171, feb 2013.

\bibitem{6736761}
E.~G. Larsson \emph{et~al.}, ``Massive mimo for next generation wireless
  systems,'' \emph{IEEE Communications Magazine}, vol.~52, no.~2, pp. 186--195,
  February 2014.

\bibitem{Harris2015}
P.~Harris \emph{et~al.}, ``{A Distributed Massive MIMO Testbed to Assess
  Real-World Performance and Feasibility},'' pp. 1--2, 2015.

\bibitem{WorldRecord}
\BIBentryALTinterwordspacing
``{Bristol and Lund set a new world record in 5G wireless spectrum
  efficiency},'' 2016. [Online]. Available:
  \url{http://www.bristol.ac.uk/news/2016/march/massive-mimo.html}
\BIBentrySTDinterwordspacing

\bibitem{WorldRecord2}
\BIBentryALTinterwordspacing
``{Bristol and Lund once again set new world record in 5G wireless spectrum
  efficiency},'' 2016. [Online]. Available:
  \url{http://www.bristol.ac.uk/news/2016/may/5g-wireless-spectrum-efficiency.html}
\BIBentrySTDinterwordspacing

\bibitem{USRP}
\BIBentryALTinterwordspacing
``{USRP-RIO 2943 Datasheet},'' 2014. [Online]. Available:
  \url{http://www.ni.com/datasheet/pdf/en/ds-538}
\BIBentrySTDinterwordspacing

\bibitem{PaulGlobecom}
P.~Harris \emph{et~al.}, ``Los throughput measurements in real-time with a
  128-antenna massive mimo testbed,'' 2016, accepted for presentation.

\bibitem{PaulSIPS}
------, ``Serving 22 users in real-time with a 128-antenna massive mimo
  testbed,'' 2016, accepted for presentation.

\bibitem{zhou2015spherical}
Z.~Zhou, X.~Gao, J.~Fang, and Z.~Chen, ``Spherical wave channel and analysis
  for large linear array in los conditions,'' in \emph{2015 IEEE Globecom
  Workshops (GC Wkshps)}.\hskip 1em plus 0.5em minus 0.4em\relax IEEE, 2015,
  pp. 1--6.

\bibitem{aulin2015benefits}
J.~Aulin, ``Benefits of variation of large scale fading across large antenna
  arrays,'' in \emph{2015 9th European Conference on Antennas and Propagation
  (EuCAP)}.\hskip 1em plus 0.5em minus 0.4em\relax IEEE, 2015, pp. 1--5.

\bibitem{athanasiadou1995ray}
G.~Athanasiadou, A.~Nix, and J.~McGeehan, ``A ray tracing algorithm for
  microcellular and indoor propagation modelling,'' in \emph{Antennas and
  Propagation, 1995., Ninth International Conference on (Conf. Publ. No. 407)},
  vol.~2.\hskip 1em plus 0.5em minus 0.4em\relax IET, 1995, pp. 231--235.

\bibitem{bohagen2006modeling}
F.~Bohagen, P.~Orten, and G.~E. Oien, ``Modeling of line-of-sight 2 {\~a} 2
  mimo channels: Spherical versus plane waves,'' in \emph{2006 IEEE 17th
  International Symposium on Personal, Indoor and Mobile Radio
  Communications}.\hskip 1em plus 0.5em minus 0.4em\relax IEEE, 2006, pp. 1--5.

\bibitem{jiang2005spherical}
J.-S. Jiang and M.~A. Ingram, ``Spherical-wave model for short-range mimo,''
  \emph{IEEE Transactions on Communications}, vol.~53, no.~9, pp. 1534--1541,
  2005.

\bibitem{narasimhan2014channel}
T.~L. Narasimhan and A.~Chockalingam, ``Channel hardening-exploiting message
  passing (chemp) receiver in large-scale mimo systems,'' \emph{IEEE Journal of
  Selected Topics in Signal Processing}, vol.~8, no.~5, pp. 847--860, 2014.

\bibitem{GSA}
\BIBentryALTinterwordspacing
``{Bristol and Lund set a new world record in 5G wireless spectrum
  efficiency}.'' [Online]. Available:
  \url{http://www.gsa.europa.eu/news/results-are-galileo-increases-accuracy-location-based-services}
\BIBentrySTDinterwordspacing

\bibitem{WangLocation}
Z.~Wang \emph{et~al.}, ``{Location-based channel estimation and pilot
  assignment for massive MIMO systems},'' \emph{ICC Workshop on 5G \& Beyond
  Enabling Technologies and Applications}, pp. 1--5, 2015.

\end{thebibliography}
%\begin{thebibliography}{1}
%\end{thebibliography}

\end{document}